\documentstyle[onecolumn]{mn}

\newif\ifAMStwofonts

\newcommand{\etal}{{et al.}~}
\newcommand{\de}{\delta}
\newcommand{\te}{\theta}

\newcommand{\Sig}{\Sigma}

\newcommand{\pa}{\partial}
\newcommand{\f}{\frac}
\newcommand{\s}{\sigma}

\newcommand{\bfr}{\bmath{r}}
\newcommand{\bfs}{\bmath{s}}

\newcommand{\bfk}{\bmath{k}}
\newcommand{\bfv}{\bmath{v}}

\newcommand{\bfu}{\bmath{u}}
\newcommand{\calB}{{\cal B}}

\newcommand{\calF}{{\cal F}}
\newcommand{\calO}{{\cal O}}

\newcommand{\calN}{{\cal N}}

\newcommand{\bc}{\begin{center}}
\newcommand{\be}{\begin{equation}}
\newcommand{\ee}{\end{equation}}
\newcommand{\ec}{\end{center}}
\newcommand{\lan}{\langle}
\newcommand{\ran}{\rangle}

\newcommand{\kms}{{${\rm km\; s^{-1}}$}}

\begin{document}

\title{Comparing the redshift-space density field to
the real-space velocity field} 

\date{}
\author[M.\ J.\ Chodorowski]
     {Micha{\l} J.\ Chodorowski\\ Copernicus Astronomical
     Center, Bartycka 18, 00--716 Warsaw, Poland\\ }

\maketitle

\begin{abstract}
I derive a nonlinear local relation between the {\em redshift-space\/}
density field and the {\em real-space\/} velocity field. The relation
accounts for radial character of redshift distortions, and it is not
restricted to the limit of the distant observer.  Direct comparisons
between the observed redshift-space density fields and the real-space
velocity fields possess all of the advantages of the conventional
redshift-space analyses, while at the same time they are free of their
disadvantages.  In particular, neither the model-dependent
reconstruction of the density field in real space is necessary, nor is
the reconstruction of the nonlinear velocity field in redshift space,
questionable because of its vorticity at second order.  The nonlinear
redshift-space velocity field is irrotational only in the distant
observer limit, and that limit is not a good approximation for shallow
catalogs of peculiar velocities currently available.  Unlike the
conventional redshift-space comparisons, the comparison proposed here
does not have to be restricted to the linear regime. Accounting for
nonlinear effects removes one of the sources of bias in the estimate
of $\beta$. Moreover, the nonlinear effects break the Omega--bias
degeneracy plaguing all analyses based on linear theory.
\end{abstract}


\section{Introduction}
\label{sec:intro}

Comparisons between density fields of galaxies and the fields of their
peculiar velocities are a powerful tool to measure the cosmological
parameter $\Omega$. This is because in the gravitational instability
paradigm the density and the velocity fields are tightly related, and
the relation between them depends on $\Omega$.

These comparisons are commonly performed in real space. A necessary
ingredient of real-space analyses is the reconstruction of the galaxy
density field in real space from the observed galaxy field, which is
given only in redshift space. While performing such a reconstruction,
one has to correct the redshifts for peculiar velocities of galaxies.
However, the amplitude of these velocities depends on $\Omega$, or in
the case of bias, on $\beta$. This is a serious drawback of real-space
comparisons: they necessarily assume the value of a parameter which is
to be subsequently estimated.

To avoid this problem, the comparisons in redshift space have been
proposed (Nusser \& Davis 1994, hereafter ND). However, the velocity
field in redshift space is irrotational only at the linear
order,\footnote{Chodorowski \& Nusser (1999) have shown that the
nonlinear redshift-space velocity field is irrotational in the distant
observer limit. However, the catalogs of peculiar velocities are not
yet deep enough for this assumption to be well satisfied (Zaroubi
\& Hoffman 1996, Szalay, Matsubara \& Landy 1998).} so that the
redshift-space analyses must be restricted to the linear regime. This
is unsatisfactory, since the derived amplitude of the density
fluctuations from current redshift surveys (e.g., Fisher \etal 1995)
and from the {\sc potent} reconstruction of density fields (Dekel
\etal 1990 and Bertschinger \etal 1990), slightly exceeds the linear
regime. For example, the density contrast in regions like the Great
Attractor or Perseus-Pisces is about unity even when smoothed over
scales of $1200$ \kms\ (Sigad \etal 1998). Future redshift surveys and
peculiar velocity catalogs are expected to provide reliable estimates
of density and velocity fields on scales where nonlinear effects are
certainly non-negligible. Those effects may lead to interesting
consequences, such as breaking the degeneracy between $\Omega$ and
bias (Dekel \etal 1993, Bernardeau \etal 1999, Chodorowski 1999;
hereafter C99), and therefore they should be accounted for.

Here I propose to compare the {\em redshift-space\/} density field
directly to the {\em real-space\/} velocity field. Such a comparison
enables one to avoid the model-dependent reconstruction of the density
field in real space. Also, the vorticity of the velocity field is no
longer a problem, because (before shell crossings) the real-space
velocity field is irrotational without any restrictions. Therefore,
the comparison proposed here combines all advantages of real-space and
redshift-space analyses, while at the same time it is free from their
disadvantages.

The present analysis is based on the work of C99, who derived a
similar relation in the distant observer limit (DOL). However, as
pointed out first by Zaroubi \& Hoffman (1996), and more recently by
Szalay, Matsubara \& Landy (1998), the DOL approximation is not well
satisfied by available catalogs of galaxy redshifts and peculiar
velocities. Here I relax this assumption, and I take fully into
account the radial character of redshift distortions. I show how the
derived relation can be inverted, i.e., how to reconstruct the
real-space velocity field from the redshift-space density field. This
is a necessary ingredient of the so-called velocity--velocity
comparisons. The results of the paper are then compared to the results
of ND.

The outline of this paper is as follows: in Section~\ref{sec:rel_ex} I
derive the exact relation between the redshift-space density and the
real-space velocity. In Section~\ref{sec:rel_app} I expand it
explicitly up to linear and second-order terms. Next, in
Section~\ref{sec:bias} I derive a similar relation between the {\em
galaxy\/} density field and the velocity field under an assumption of
a nonlinear but local bias. In Section~\ref{sec:Omega_bias} I show how
such a relation can be used to break the $\Omega$--bias degeneracy in
velocity--velocity comparisons. Summary and conclusions are in
Section~\ref{sec:summ}.

\section{Exact relation}
\label{sec:rel_ex}

The transformation from the real space coordinate, $\bfr$, to the
redshift space coordinate, $\bfs$, is (Kaiser 1987; hereafter K87)

\be 
\bfs = \bfr + v_r \hat{\bfr} \,,
\label{eq:rtos}
\ee
where $v_r(\bfr) = \bfv \cdot \hat{\bfr}$; $\bfv(\bfr)$ is the
real-space velocity field, and velocities are measured relative to the
Local Group. The relation between the redshift-space mass density
field, $\de_s$, and the real-space velocity field can be obtained from
the equation

\be
\de_{s}(\bfs) = \sum_{n = 0}^{\infty} \f{(-1)^n}{n!}
\left. \left\{ \f{1}{r^2} \f{\pa^n}{\pa r^n} 
\left[ r^2 v_r^n 
\left( \f{\phi(r)}{\phi[s(r)]} (\de + 1) - J \right) \right] 
\right\} \right|_{\tilde{\bfr} = \bfs} \,.
\label{eq:gen}
\ee
Here, $\phi$ is the selection function, $\de(\bfr)$ is the real-space
density field, and $J(\bfr)$ is the Jacobian of the
transformation~(\ref{eq:rtos}),

\be
J(\bfr) = (1 + v_r/r)^2 (1 + v_r') \,,
\label{eq:jac}
\ee
where $' \equiv \pa/\pa r$.

Equation~(\ref{eq:gen}) follows from the continuity equation; its
derivation is outlined in Appendix~\ref{app:gen}. The equation is
exact; however it is strictly valid only in the absence of
triple-value zones in redshift space. It is a generalization of
equation~(A6) of C99, which was derived in the DOL.

The relation between the redshift-space density and the real-space
velocity follows from equation~(\ref{eq:gen}) when the
density--velocity relation (DVR) in real space is used. The DVR in
real space is well known, it has been extensively studied both
analytically and numerically (Nusser \etal 1991, Bernardeau 1992,
Gramann 1993, Mancinelli \etal 1994, Mancinelli \& Yahil 1995,
Chodorowski 1997, Chodorowski \& {\L}okas 1997, Chodorowski \etal
1998, Bernardeau \etal 1999, Ganon \etal 2000, Kudlicki \etal
2000). While a number of accurate formulas has been proposed, for the
purpose of this Section it is sufficient that the real-space density
is a well-known nonlinear function of the first-order velocity
derivatives, with an explicit dependence on $\Omega$,

\be
\de(\bfr) = \calF[\pa v_i/\pa r_j(\bfr),\Omega] \,.
\label{eq:dvr_real}
\ee
If the selection function is given by a power law, $\phi \propto s^{-p}$, then

\be
\f{\phi(r)}{\phi[s(r)]} = (1 + v_r/r)^p \,.
\label{eq:sel}
\ee
If not, the expression $\phi(r)/\phi(s)$ should be expanded
explicitly. In any case, it will be a function of $v_r/r$. Therefore,
both $\phi(r)/\phi(s)$, $\de$ and $J$ are functions of the real-space
velocity field and its derivatives, and equation~(\ref{eq:gen})
becomes a relation between the redshift-space density and the
real-space velocity.

The key point of equation~(\ref{eq:gen}) is that it relates the
redshift-space density at a point $\bfs$ to an expression involving
the real-space velocity field evaluated at $\tilde{\bfr} =
\bfs$. ($\tilde{\bfr}$ is a real-space point, in general different
from $\bfr$, which is related to $\bfs$ by eq.~\ref{eq:rtos}.) This
has been achieved by effectively expanding real-space quantities,
appearing in the continuity equation as functions of $\bfr$, around
$\tilde{\bfr} = \bfs$ (for details see C99 and
Appendix~\ref{app:gen}). This is very convenient, since now we can
treat the two fields as if they were given in the same coordinate
system. In other words, we can compare the redshift-space density at a
redshift, say, $3000$ \kms, directly to the real-space velocity at a
real-space position $3000$ \kms.

Relation~(\ref{eq:gen}) involves an infinite series in velocity. If
the density--velocity comparison is performed on scales large enough
so that they are only weakly nonlinear, then this series can be
truncated. Truncation at order $q$ means that the terms of the order
$q+2$ in velocity or higher are neglected. The truncation procedure
should be done with care, because nonlinear effects are stronger in
redshift space than in real space (e.g., Taylor \& Hamilton 1996,
Scoccimarro, Couchman \& Frieman 1998, this paper). For example, the
formalism presented here fails for triple-value zones (regions with
shell crossings in redshift space). However, these regions or
structures in real space may be just after turning around, quite a
mildly nonlinear condition. Thus far, all density--velocity
comparisons in redshift space have been restricted to linear regime
(e.g., ND, Fisher \etal 1995). In the next section I will derive an
explicit expression for the relation~(\ref{eq:gen}) up to second-order
terms. Moreover, it will become clear how to extend the derivation to
still higher order, if necessary.

\section{Approximate relation}
\label{sec:rel_app}

\subsection{Linear relation}
To derive the linear relation, we linearize the ratio of the selection
functions, the real-space DVR, and the Jacobian. We have

\be
\f{\phi(r)}{\phi[s(r)]} \simeq 1 - \f{d \ln \phi}{d \ln r} \f{v_r}{r} \,,
\label{eq:sel_1}
\ee

\be 
\de \simeq - f^{-1} \nabla_{\bfr} \cdot \bfv \,,
\label{eq:dvr_1}
\ee
and

\be
J \simeq 1 + v_r' + 2 v_r/r \,.
\label{eq:jac_1}
\ee

Hence, 

\be
\f{\phi(r)}{\phi[s(r)]} (\de + 1) - J \simeq
- f^{-1} \nabla_{\bfr} \cdot \bfv - v_r' - 
\left(2 + \f{d \ln \phi}{d \ln r} \right) \f{v_r}{r} \,,
\label{eq:expr}
\ee
and we see that in the series~(\ref{eq:gen}), only the term with $n =
0$ contributes to linear order. Therefore,

\be
\de_s(\bfs) = 
\left. \left[ - f^{-1} \nabla_{\bfr} \cdot \bfv - v_r' - 
\left(2 + \f{d \ln \phi}{d \ln r} \right) \f{v_r}{r} 
\right] \right|_{\tilde{\bfr} = \bfs} \,. 
\label{eq:de_s_1}
\ee

This equation is equivalent to equation~(3.3) of K87. It also coincides
with equation~(6) of ND. However, the equation of ND describes the
relation between the redshift-space density and the {\em
redshift-space\/} velocity, $\bfu(\bfs)$, and all derivatives are
taken there with respect to $\bfs$. The two equations coincide,
because

\be
\pa u_i/\pa s_j = \pa v_i/\pa r_j + \calO\left( v^2 \right) \,.
\ee
Clearly, the equations would be different at the second order.

\subsection{Second-order relation}
In the series~(\ref{eq:gen}), only the terms with $n=0$ and $n=1$ generate the
terms up to quadratic in velocity. To second order,

\be
\f{\phi(r)}{\phi[s(r)]} \simeq 1 -  D_1 v_r/r + 
\left(D_1^2 - D_2\right) (v_r/r)^2 \,,
\label{eq:sel_2}
\ee
where 

\be 
D_1 = \f{d \ln \phi}{d \ln r} \,,
\label{eq:d1}
\ee
and 

\be 
D_2 = \f{\phi'' r^2}{2 \phi} \,.
\label{eq:d2}
\ee
Next, 

\be
J \simeq 1 + v_r' + 2 v_r/r + 2 v_r' v_r/r + (v_r/r)^2 \,.
\label{eq:jac_2}
\ee
Finally, the real-space density is up to second order a local function of the
two velocity scalars: the expansion (the divergence), $\te$, and the shear,
$\Sig$. Specifically (Chodorowski 1997; see also Gramann 1993, Catelan \etal
1995, Mancinelli \& Yahil 1995)

\be
\de(\bfr) = - f^{-1} \te(\bfr) +
{\textstyle \f{4}{21}} f^{-2} 
\left[\te^2(\bfr) - {\textstyle \f{3}{2}} \Sig^2(\bfr) \right] \,.
\label{eq:del_2}
\ee
Here,
\be
\Sig^2 \equiv \Sig_{ij} \Sig_{ij} \,,
\label{eq:shear_scalar}
\ee
\be
\Sig_{ij} \equiv {\textstyle \f{1}{2}}
\left(\pa v_i/\pa r_j + \pa v_j/\pa r_i \right)
- {\textstyle \f{1}{3}} \de^{K}_{ij} \te
\label{eq:shear_tensor}
\,,
\ee
\be
\te \equiv \nabla_{\bfr} \cdot \bfv
\,,
\label{eq:div}
\ee
and I have neglected the weak $\Omega$-dependence. The symbol $\de^{K}_{ij}$
denotes the Kronecker delta.

From equations~(\ref{eq:sel_2}), (\ref{eq:jac_2}),
and~(\ref{eq:del_2}) the contribution from the term $n=0$ in the
series~(\ref{eq:gen}) can be obtained. As for the term $n=1$, there is
an extra factor $v_r$ in front of the expression in round parentheses
($\phi(r)/\phi(s) \ldots$). Therefore, it is sufficient to adopt for
this expression its linear approximation~(\ref{eq:expr}). After a
straightforward but rather lengthy algebra we obtain
\begin{eqnarray}
\de_s(\bfs) 
&=& 
\left\{
- f^{-1} \te - v_r' - (2 + D_1) v_r/r 
+ \left[v_r(f^{-1} \te + v_r')\right]'
+ {\textstyle \f{4}{21}} f^{-2}
\left(\te^2 - {\textstyle \f{3}{2}} \Sig^2\right) 
\right.
\nonumber \\
&~&
\left. \left.
+ (2 + D_1) \left(f^{-1} \te + 2 v_r'\right)^{\vphantom{'}} v_r/r 
+ \left(1 + D_1 + D_1^2 - D_2 \right) (v_r/r)^2
\right\} \right|_{\tilde{\bfr} = \bfs} \,. 
\label{eq:de_s_2}
\end{eqnarray}

In the DOL, $v_r/r \to 0$ and lines-of-sight are taken at a fixed
direction, so $v_r \to v_z$ and $\pa/\pa r \to \pa/\pa z$, and the
above formula reduces to equation~(13) of C99.
 
K87 derived a relation between the redshift-space density and the
real-space velocity only at the linear order. To derive a related
relation, between the redshift-space density and the {\em
redshift}-space velocity, ND applied the Zel'dovich approximation.
However, they subsequently restricted their analysis again to the
linear regime. The reason they did so was that the redshift-space
velocity field is rotational at second order, so that it cannot be
reconstructed from radial components of velocities solely. The
analysis presented here is free of this problem since we relate the
redshift-space density field directly to the real-space velocity
field, which is irrotational without any restrictions (before shell
crossings in real space). 

Applying linear theory in density--velocity comparisons leads to a
bias in the inferred value of $\Omega$. We can estimate the amplitude
of this bias using equation~(\ref{eq:de_s_2}). For simplicity, adopt
the DOL and consider a spherical top-hat overdensity. Then

\be
\de_{s} =  - \left(1 + \f{f}{3} \right) f^{-1} \te 
+ \left(\f{4}{21} + \f{f}{3} + \f{f^2}{9} \right) f^{-2} \te^2
\label{eq:tophat}
\ee 
(cf.\ eq.~17 of C99). Assuming the second term on the right-hand side of
the above equation to be a small correction it is straigthforward to
show that the relative systematic error of $\Omega$ is 

\be
\f{\Delta \Omega}{\Omega} \simeq \f{5}{3} 
\left(\f{4}{21} \Omega^{-0.6} + \f{1}{3} + \f{1}{9} \Omega^{0.6} \right) 
\te \,.
\label{eq:error}
\ee 
In the Mark III catalog, the r.m.s.\ value of $\te$, when smoothed
over scales of $1200$ \kms, is $\sim 0.3$ (Sigad \etal 1998). Hence,
the relative bias of the linear estimate of $\Omega$ in {\em redshift
space\/} is about $40$\% for $\Omega = 0.3$, and about $30$\% for
$\Omega = 1.0$. This may be compared to the corresponding bias of the
linear estimate of $\Omega$ in {\em real space,\/} about $20$ and
$10$\% respectively. (It is obtained by setting the second and the
third terms in parentheses in equation~\ref{eq:error} equal to zero.)
Thus, the nonlinear effects in redshift space are stronger than in
real space. Obviously, the bias will be respectively greater for
smaller smoothing scales.

Equation~(\ref{eq:de_s_2}) can be used to reconstruct the real-space velocity
field from the associated redshift-space density field. Since the real-space
velocity field is irrotational, it can be described as a gradient of the
velocity potential,

\be
\bfv(\bfr) = - \nabla_{\bfr} \Phi_v \,.
\label{eq:vel_pot}
\ee
Equation~(\ref{eq:de_s_2}) reduces then to a nonlinear differential
equation for the velocity potential. As long as the nonlinearities
in~(\ref{eq:de_s_2}) are weak, an iterative method of solution can be
applied. First, we solve the linear part of~(\ref{eq:de_s_2}), and the
method of ND (the decomposition in spherical harmonics) is fully
applicable. Next, we find the second-order solution by solving again
the linear equation, with the source term resulting from the density
modified by nonlinear contributions approximated by first-order
solutions. Specifically,

\be
f^{-1} \Delta_{\bfr} \Phi_v^{(2)} + 
\f{\pa^2 \Phi_v^{(2)}}{\pa r^2} + 
\f{2 + D_1}{r} \f{\pa \Phi_v^{(2)}}{\pa r} = \de_s(\tilde{\bfs} = \bfr) - 
\calN_2\left[\Phi_v^{(1)}(\bfr)\right]
\,,
\label{eq:pot_2}
\ee
where $\calN_2$ is a sum of all terms quadratic in velocity in
equation~(\ref{eq:de_s_2}), expressed as functions of the potential. 

In this way, from the redshift-space density field we can reconstruct
the real-space velocity potential, and hence the real-space velocity
field. The latter can then be compared to measured radial velocities
of galaxies. Note that this is a velocity--velocity comparison; the
application of the `redshift-real' DVR to density--density comparisons
has been discussed in C99.

\section{Galaxy density versus velocity}
\label{sec:bias}

Galaxies are good tracers of the velocity field induced by the mass
distribution. On the other hand, they are biased tracers of the mass
distribution itself (see theoretical arguments of Kaiser 1984, Davis
\etal 1985, Bardeen \etal 1986, Dekel \& Silk 1986, Cen \& Ostriker
1992, Kauffmann, Nusser \& Steinmetz 1997 and observational evidence
by Davis \& Geller 1976, Dressler 1980, Giovanelli, Haynes \&
Chincarini 1986, Santiago \& Strauss 1992, Loveday \etal 1996, Hermit
\etal 1996, Guzzo \etal 1997, Giavalisco \etal 1998). In this section
I will derive a relation between the redshift-space {\em galaxy\/}
density field and the real-space velocity field under the assumption
of a nonlinear but local bias. This is only a toy model for bias
because there are good reasons to believe that bias is in fact
somewhat stochastic (Dekel \& Lahav 1998, Pen 1998, Tegmark \& Peebles
1998, Blanton \etal 1998, Blanton \etal 1999, Tegmark \& Bromley
1998). However, it allows for a number of important conclusions to be
drawn. 

Equation~(\ref{eq:gen}) implicitly assumes no bias between the
distribution of galaxies and mass, so that the densities appearing in
it are in fact the {\em galaxy\/} densities $\de_{s}^{\rm (g)}$ and
$\de^{\rm (g)}$ (redshift- and real-space galaxy density contrasts
respectively).  Hence, a more general form of equation~(\ref{eq:gen})
is

\be
\de_{s}^{\rm (g)}(\bfs) = \sum_{n = 0}^{\infty} \f{(-1)^n}{n!}
\left. \left\{ \f{1}{r^2} \f{\pa^n}{\pa r^n} 
\left[ r^2 v_r^n \left( \f{\phi(r)}{\phi[s(r)]} 
\left[\de^{\rm (g)} + 1\right] - J \right) \right] 
\right\} \right|_{\tilde{\bfr} = \bfs} .
\label{eq:gen_gal}
\ee 
In a local bias model, the real-space galaxy density contrast is
assumed to be in general a nonlinear function of the mass density
contrast (Fry \& Gazta\~naga 1993; see also Juszkiewicz \etal 1995),

\be
\de^{\rm (g)}(\bfr) = \calB\left[\de(\bfr)\right]
\,. 
\label{eq:bias}
\ee
Given the real-space DVR~(\ref{eq:dvr_real}),
equation~(\ref{eq:gen_gal}) yields

\be
\de_{s}^{\rm (g)}(\bfs) = \sum_{n = 0}^{\infty} \f{(-1)^n}{n!}
\left. \left\{ \f{1}{r^2} \f{\pa^n}{\pa r^n} 
\left[ r^2 v_r^n \left( \f{\phi(r)}{\phi[s(r)]} 
\left[\calB \circ \calF(\pa v_i/\pa r_j) + 1\right] - J \right) \right] 
\right\} \right|_{\tilde{\bfr} = \bfs} \,.
\label{eq:dv_bias}
\ee 

Up to second order, $\calF$ is given by expression~(\ref{eq:del_2}) and

\be
\calB\left[\de(\bfr)\right] = b \de(\bfr) + {\textstyle \f{1}{2}} b_2 
\left[\de^2(\bfr) - \s_{\de}^2\right]  
\,.
\label{eq:bias_2}
\ee 
Here, $b$ and $b_2$ are respectively the linear and nonlinear
(second-order) bias parameters. The term $\s_{\de}^2 \equiv \lan \de^2
\ran$ ensures that the mean value of $\de^{\rm (g)}$ vanishes, as
required. Combining~(\ref{eq:bias_2}) and~(\ref{eq:dv_bias}), and
truncating the series at second-order terms we obtain\footnote{For
simplicity, I have approximated the quantity $\te^2 - \s_{\te}^2$ by
$\te^2 - (3/2) \Sig^2$. This is legitimate, since the mean value of
the shear scalar given the velocity divergence, $\lan
\Sig^2\ran|_{\te}$, is up to second order equal to $\lan \Sig^2\ran =
(2/3) \s_{\te}^2$ (Chodorowski 1997).}
\begin{eqnarray}
\de_s^{\rm (g)}(\bfs) 
&=& 
\left\{
- \beta^{-1} \te - v_r' - (2 + D_1) v_r/r 
+ \left[v_r(\beta^{-1} \te + v_r')\right]'
+ \left(\f{4}{21 b} + \f{b_2}{2 b^2}\right) \beta^{-2}
\left(\te^2 - {\textstyle \f{3}{2}} \Sig^2\right) 
\right.
\nonumber \\
&~&
\left. 
+ (2 + D_1) \left(\beta^{-1} \te + 2 v_r'\right) v_r/r 
+ \left(1 + D_1 + D_1^2 - D_2 \right) (v_r/r)^2
\Biggr\} \right|_{\tilde{\bfr} = \bfs} \,. 
\label{eq:dv_bias_2}
\end{eqnarray}

Expressing velocity in terms of the velocity potential yields

\be
\beta^{-1} \Delta_{\bfr} \Phi_v^{(2)} + 
\f{\pa^2 \Phi_v^{(2)}}{\pa r^2} + 
\f{2 + D_1}{r} \f{\pa \Phi_v^{(2)}}{\pa r} = 
\de_s^{\rm (g)}(\tilde{\bfs} = \bfr) - 
\calN_2\left[\Phi_v^{(1)}(\bfr),\beta,\beta_2\right]
\,,
\label{eq:pot_bias}
\ee
where
\be
\beta = f(\Omega)/b 
\label{eq:beta}
\ee
and
\be
\beta_2 = \left(\f{4}{21 b} + \f{b_2}{2 b^2}\right)^{-1} \beta^2
\,.
\label{eq:beta_2}
\ee

\section{Measuring $\bmath{\lowercase{\Omega}}$ and bias
separately$\,^{\textstyle \ddag}$}

\label{sec:Omega_bias}

Based\addtocounter{footnote}{1}\footnotetext{This Section is similar
to Section~5 of C99. I include it here for the completeness of the
present paper.} on~(\ref{eq:pot_bias}), we can reconstruct the
real-space velocity field from the associated redshift-space galaxy
density field. Comparison of the predicted velocity field to the
observed one will yield the best-fit values of the parameters $\beta$
and $\beta_2$. They are a combination of three physical parameters:
$\Omega$, $b$, and $b_2$. Therefore, we have two constraints for three
parameters, so we need an additional constraint on them. As this
constraint one can adopt the large-scale galaxy density skewness,
which involves $b$ and $b_2$ (B99).

Galaxy skewness can be measured only in redshift space. Therefore,
we need a theoretical relation between the redshift-space galaxy
density skewness $S_{3 s}^{\rm (g)}$ (which we can measure) and the
redshift-space mass density skewness $S_{3 s}$ (which we can
compute). Approximately this relation is (see C99 for details),

\be
S_{3 s}^{\rm (g)} = \f{S_{3 s}}{b} + 3 \f{b_2}{b^2} 
\,.
\label{eq:S_3}
\ee

The redshift-space mass density skewness was computed for scale-free
power spectra by Hivon \etal (1995). They found that it depends only
very weakly on $\Omega$ and that it can be well approximated by the
real-space skewness, except in the case of $\Omega$ approaching unity
{\em and\/} the spectral index $n \ga 0$. For more realistic power
spectra the calculation can be done in an analogous way, and the result
is also expected to be well approximated by the corresponding
real-space value.

The redshift-space galaxy density skewness of the {\it IRAS\/} density
field was measured directly by Bouchet \etal (1993). Kim \& Strauss
(1998) pointed out that sparse sampling of the {\it IRAS\/} galaxies,
coupled with the moments method used by Bouchet et al., makes the
skewness estimates biased low. However, they proposed a method to
measure the skewness by fitting the Edgeworth expansion to the galaxy
count probability distribution function around its maximum, properly
accounting for the shot noise. Using mock catalogs, Kim \& Strauss
showed that this estimate of skewness is robust to sparse sampling;
they then used this method to measure the skewness of the {\it IRAS\/}
density field. Hence, we know the value of the skewness of the
distribution of the {\it IRAS\/} galaxies (and we know an
effective means to measure it in other galaxy catalogs).

Equations~(\ref{eq:beta})--(\ref{eq:S_3}) are three independent
constraints for the parameters $\Omega$, $b$ and $b_2$, so that they
can be used to measure $\Omega$ and bias separately. The additional
constraint on $\Omega$ and bias (the skewness) is to be inferred from
the density field alone, making any additional observations
unnecessary. In other words, there is enough information in the
density field and the corresponding velocity field to break the
$\Omega$--bias degeneracy.

\section{Summary}
\label{sec:summ}

I have derived a nonlinear local relation between the {\em
redshift-space\/} galaxy density field and the {\em real-space\/}
velocity field. A direct comparison of the redshift-space density
directly to the real-space velocity makes the model-dependent
reconstruction of the density field in real space unnecessary.
Furthermore, the comparison need not to be restricted to the linear
regime because the nonlinear real-space velocity field, unlike the
redshift-space one, is irrotational.

The analysis presented here is an extension of the work of C99, who
performed a similar analysis in the distant observer limit. However,
available catalogs of peculiar velocities are too shallow for this
limit to be approached (Zaroubi \& Hoffman 1996, Szalay, Matsubara \&
Landy 1998). Here I have relaxed this approximation, and I have taken
the radial nature of redshift distortions fully into account. C99
showed how to apply the `redshift-real' density--velocity relation to
the so-called density--density comparisons. Here I have shown how to
apply it to velocity--velocity comparisons as well. Namely, I have
derived an equation enabling one to reconstruct the real-space
velocity field from the redshift-space galaxy density field. I have
explicitly derived the density--velocity relation up to linear and
second-order terms, but the derivation can be straightforwardly
extended to any desired order using the exact
equation~(\ref{eq:dv_bias}). I have compared the linear relation to
that of K87 and of ND and found them to be in agreement.

Applying the second-order relation derived here, or its higher-order
extension, to density--velocity comparisons will remove one of the
sources of bias (e.g., the nonlinear effects) in the estimate of
$\beta$, or $\Omega$. I have shown this bias to be substantial in the
linear theory. For the fields smoothed over scales as large as $1200$
\kms, the relative systematic error of the linear estimate of $\Omega$
is $30$--$40$\% (depending on the value of $\Omega$). Moreover,
combining the nonlinear relation with an additional measurement (e.g.,
the skewness of the galaxy density field) breaks the degeneracy
between $\Omega$ and the galaxy bias.

The ultimate goal of density--velocity comparisons is to estimate the
value of $\Omega$ separately from the galaxy bias. The method proposed
here offers such a possibility, and it will be developped in
forthcoming papers. Issues remaining to be addressed are the spatial
scales at which a perturbative approach in redshift space can be
applied, the stochastic character of bias, and effects of field
smoothing. Kudlicki \& Chodorowski (in preparation) study the first
issue. Taruya \& Soda (1998) derived a mildly nonlinear galaxy
density--mass density relation in a stochastic bias model. The
calculation of the galaxy density--velocity relation in this model
will be to some extent similar. Field smoothing affects the real-space
density--velocity relation in a very weak way (Chodorowski \& {\L}okas
1997, Chodorowski \etal 1998, Bernardeau \etal 1999). The effects of
smoothing are somewhat stronger in redshift space (C99). In any case,
they should be studied individually for any particular comparison,
because of different smoothing schemes used. For example, in the
analysis of ND, both the density and the velocity fields are smoothed,
while in the {\sc velmod} analysis of Willick \& Strauss (1998), only
the {\it IRAS\/} density field is smoothed. In accounting for
smoothing, the method of C99 may be applied.

\section*{Acknowledgments}
I thank Micha{\l} R\'o\.zyczka for useful comments on the text. This
research has been supported in part by the Polish State Committee for
Scientific Research grants No.~2.P03D.008.13 and 2.P03D.004.13.

\appendix
\section{Derivation of
equation~(\lowercase{\ref{eq:gen}})}
\label{app:gen}

Here I outline the derivation of the exact equation~(\ref{eq:gen}), expressing
the redshift-space density field at $\bfs$ in terms of the real-space
quantities at $\tilde{\bfr} = \bfs$. The derivation follows to some extent the
lines of Appendix~A of C99, where a similar equation has been derived in the
DOL.

From the conservation of the number of galaxies we have 

\be
\de_{s}(\bfs) = \f{\phi(r)}{\phi[s(r)]} J^{-1}(\bfr) [\de(\bfr) + 1] - 1 
\,,
\label{eq:cons}
\ee
where $\phi$ is the selection function and $J$ is the Jacobian of the mapping
from real to redshift space. The Fourier transform of the redshift-space
density contrast, using the mapping~(\ref{eq:rtos}), is

\be
\de_{s}(\bfk) \equiv \int \f{{\rm d}^3 s}{(2 \pi)^3} {\rm
e}^{-i \bfk \cdot \bfs} \de_{s}(\bfs) = 
\int \f{{\rm d}^3 r}{(2 \pi)^3} 
{\rm e}^{-i \bfk \cdot \bfr} {\rm e}^{- i k_r v_r(\bfr)} 
\left\{ \f{\phi(r)}{\phi[s(r)]} [\de(\bfr) + 1] - J(\bfr) \right\}
\label{eq:ds(k)}
\ee
(cf.\ eq.~4 of Scoccimarro \etal 1998 in the DOL). 
The inverse Fourier transform of the above equation is

\be 
\de_{s}(\bfs) \equiv \int {\rm d}^3 k \, {\rm e}^{i
\bfk \cdot \bfs} \de_{s}(\bfk) = 
\int \f{{\rm d}^3 r \, {\rm d}^3 k}{(2 \pi)^3} 
\left\{ \f{\phi(r)}{\phi[s(r)]} [\de(\bfr) + 1] - J(\bfr) \right\}
{\rm e}^{- i k_r v_r(\bfr)} {\rm e}^{i \bfk \cdot (\bfs - \bfr)} \,.
\label{eq:ds(s)}
\ee
Here, I have changed the order of integration. Expanding the first exponent of
the integrand we have

\be 
{\rm e}^{- i k_r v_r(\bfr)} {\rm e}^{i \bfk \cdot
(\bfs - \bfr)} =
\sum_{n=0}^\infty \f{v_r^n(\bfr)}{n!} (- i k_r)^n 
{\rm e}^{i \bfk \cdot (\bfs - \bfr)} =
\sum_{n=0}^\infty \f{v_r^n(\bfr)}{n!}
\f{\partial^n}{\partial r^n} {\rm e}^{i \bfk \cdot (\bfs - \bfr)} \,,
\label{eq:expan}
\ee
hence

\be
\de_{s}(\bfs) =
\sum_{n=0}^\infty \f{1}{n!} \int {\rm d}^3 r \, v_r^n(\bfr)
\left\{ \f{\phi(r)}{\phi[s(r)]} [\de(\bfr) + 1] - J(\bfr) \right\}
\f{\partial^n}{\partial r^n}
\int \f{{\rm d}^3 k}{(2 \pi)^3} {\rm e}^{i \bfk \cdot (\bfs - \bfr)}
\,.
\label{eq:ds(s)2}
\ee
The integral over $k$ yields the Dirac delta
distribution, $\de_{\rm D}(\bfs - \bfr)$. Now the following lemma will
be useful:

\be
\int {\rm d}^3 r \, g(\bfr) \f{\pa^n}{\pa r^n} \de_{\rm D}(\tilde{\bfr} -
\bfr) = \left. (-1)^n 
\left\{\f{1}{r^2} \f{\pa^n}{\pa r^n} \left[r^2 g(\bfr) \right]\right\} 
\right|_{\tilde{\bfr}} \,,
\label{eq:lemma}
\ee
which can be easily proven by integrating by parts. Using this lemma
in equation~(\ref{eq:ds(s)2}) yields equation~(\ref{eq:gen}).


\begin{thebibliography}{}
\bibitem[]{bar86} Bardeen J., Bond J. R., Kaiser N., Szalay A., 1986, 
ApJ, 304, 15
\bibitem[]{b92} Bernardeau F., 1992, ApJ, 390, L61
\bibitem[]{ber99} Bernardeau F., Chodorowski M. J., {\L}okas E. L., 
Stompor R. Kudlicki A., 1999, MNRAS, 309, 543 
\bibitem[]{potent2} Bertschinger E., Dekel A., Faber S. M., Dressler
A., Burstein D., 1990, ApJ, 364, 370
\bibitem[]{bla98} Blanton M., Cen R., Ostriker J. P., Strauss M. A., 
1998, astro-ph/9807029
\bibitem[]{bla99} Blanton M., Cen R., Ostriker J. P., Strauss M. A., 
1999, astro-ph/9903165
\bibitem[]{bou93} Bouchet F. R., Strauss M. A., Davis M., Fisher K. B., 
Yahil A., Huchra J. P., 1993, ApJ, 417, 36
\bibitem[]{cat95} Catelan P., Lucchin F., Matarrese S., Moscardini L., 
1995, MNRAS, 276, 39
\bibitem[]{co92} Cen R., Ostriker J. P., 1992, ApJ, 399, L113
\bibitem[]{ch97} Chodorowski M. J., 1997, MNRAS, 292, 695
\bibitem[]{c99} Chodorowski M. J., 1999, astro-ph/9812075 (C99) 
\bibitem[]{cl97} Chodorowski M. J., {\L}okas E. L., 1997, MNRAS, 287, 591
\bibitem[]{cn99} Chodorowski M. J., Nusser A., 1999, astro-ph/9905328 
\bibitem[]{chod98} Chodorowski M. J., {\L}okas E. L., Pollo A., Nusser A., 
1998, MNRAS, 300, 1027
\bibitem[]{dg76} Davis M., Geller M. J., 1976, ApJ, 208, 13 
\bibitem[]{dav85} Davis M., Efstathiou G., Frenk C. S., White S. D. M., 
1985, ApJ, 292, 371
\bibitem[]{dl98} Dekel A., Lahav O., 1998, astro-ph/9806193
\bibitem[]{ds86} Dekel A., Silk J., 1986, ApJ, 303, 39
\bibitem[]{potent1} Dekel A., Bertschinger E., Faber S. M., 1990, ApJ,
364, 349
\bibitem[]{dek93} Dekel A., Bertschinger E., Yahil A., Strauss M.,
  Davis M., Huchra J., 1993, ApJ, 412, 1
\bibitem[]{dress80} Dressler A., 1980, ApJ, 236, 351
\bibitem[]{fish95} Fisher K. B., Lahav O., Hoffman Y., Lynden-Bell D.,
Zaroubi S., 1995, MNRAS, 272, 885
\bibitem[]{fg93} Fry J. N., Gazta\~naga E., 1993, ApJ, 413, 447
\bibitem[]{gan99} Ganon G., Dekel A., Mancinelli P. J., Yahil A., 2000, 
in preparation
\bibitem[]{gia98} Giavalisco M. et al., 1998, astro-ph/9802318
\bibitem[]{ghc86} Giovanelli R., Haynes M. P., Chincarini G. L., 1986, 
ApJ, 300, 77
\bibitem[]{gramann} Gramann M., 1993, ApJ, 405, L47
\bibitem[]{guz97} Guzzo L., Strauss M. A., Fisher K. B., Giovanelli R., 
Haynes M. P., 1997, ApJ, 489, 37
\bibitem[]{herm96} Hermit S., Santiago B. X., Lahav O., Strauss M. A., 
Davis M., Dressler A., Huchra J. P., 1996, MNRAS, 283, 709
\bibitem[]{hetal95} Hivon E., Bouchet F. R., Colombi S., Juszkiewicz R., 
1995, A\&A, 298, 643
\bibitem[]{jusz} Juszkiewicz R., Weinberg D. H., Amsterdamski P., 
Chodorowski M. J., Bouchet F. R., 1995, ApJ, 442, 39
\bibitem[]{kai84} Kaiser N., 1984, ApJL, 284, L9
\bibitem[]{kai87} Kaiser N., 1987, MNRAS, 227, 1 (K87)
\bibitem[]{kns} Kauffman G., Nusser A., Steinmetz M., 1997, MNRAS, 286, 795
\bibitem[]{ks98} Kim R. S., Strauss M. A., 1998, ApJ, 493, 39 
\bibitem[]{kud00} Kudlicki A., Chodorowski M. J., Plewa T.,
R\'o\.zyczka M., 2000, MNRAS, in press
\bibitem[]{lov96} Loveday J., Efstathiou G., Maddox S. J., Peterson B. A., 
1996, ApJ, 468, 1
\bibitem[]{my95} Mancinelli P. J., Yahil A., 1995, ApJ, 452, 75
\bibitem[]{mygd} Mancinelli P. J., Yahil A., Ganon G., Dekel A., 1994,
    in Bouchet F. R., Lachi\`{e}ze-Rey M., eds, Proc. 9th IAP
    Astrophysics Meeting, Cosmic Velocity Fields. Editions
    Fronti\`{e}res, Gif-sur-Yvette, p. 215
\bibitem[]{nd94} Nusser A., Davis M., 1994, ApJ, 421, L1 (ND)
\bibitem[]{ndbb} Nusser A., Dekel A., Bertschinger E., Blumenthal G. R., 
1991, ApJ, 379, 6
\bibitem[]{pen} Pen U., 1998, ApJ, 504, 601
\bibitem[]{ss92} Santiago B. X., Strauss M. A., 1992, ApJ, 387, 9
\bibitem[]{scf98} Scoccimarro R., Couchman H. M. P., Frieman J. A., 1998, 
astro-ph/9808305
\bibitem[]{sig98} Sigad Y., Eldar A., Dekel A., Strauss M. A., 
Yahil A., 1998, ApJ, 495, 516
\bibitem[]{szml98} Szalay A. S., Matsubara T., Landy S. D., 1998, ApJ, 498,
L1
\bibitem[]{ts98} Taruya A., Soda J., 1998, astro-ph/9809204
\bibitem[]{th96} Taylor A. N., Hamilton A. J. S., 1996, MNRAS, 282, 767
\bibitem[]{tb98} Tegmark M., Bromley B. C., 1998, astro-ph/9809324 
\bibitem[]{tp98} Tegmark M., Peebles P. J. E., 1998, ApJ, 500, L79
\bibitem[]{ws} Willick J. A., Strauss M. A., 1998, astro-ph/9801307 
\bibitem[]{zh96} Zaroubi S., Hoffman Y., 1996, ApJ, 462, 25

\end{thebibliography}
\end{document}